\documentclass[aps,prb,reprint,superscriptaddress]{revtex4-1}
\usepackage{amsmath}
\usepackage{graphicx}
\usepackage{units}  
\usepackage{xspace}
\usepackage{subfigure}
\usepackage{hyperref}

\newcommand{\ve}{\varepsilon}
\newcommand{\mb}{\mathbf}
\newcommand{\tb}{\textbf}
\newcommand{\beq}{\begin{equation}}
\newcommand{\eeq}{\end{equation}}
\newcommand{\bea}{\begin{eqnarray}}
\newcommand{\eea}{\end{eqnarray}}

\newcommand{\mK}{\mathbf{K}}

\usepackage[usenames]{color}

\begin{document}

\bibliographystyle{apsrev}
 
\title{Quasi-flat-band physics in a two-leg ladder model and its relation to magic-angle twisted bilayer graphene}

\date{\today}

\author{Yixuan Huang}  
\affiliation{Texas Center for Superconductivity and Department of Physics, University of Houston, Houston, TX 77204, USA} 
\author{Pavan Hosur}   
\affiliation{Texas Center for Superconductivity and Department of Physics, University of Houston, Houston, TX 77204, USA}   
\author{Hridis K. Pal}
   \email{hridis.pal@iitb.ac.in}
\affiliation{Department of Physics, Indian Institute of Technology Bombay, Powai, Mumbai 400076, India}

\begin{abstract}
We study the single- and many-particle properties of a two-leg ladder model threaded by a flux with the legs coupled by a spatially varying term. Although \emph{a priori} unrelated to twisted bilayer graphene (TBG), the model is found to have striking similarities: A quasi-flat low-energy band emerges with characteristics similar to that of magic angle TBG. We study the effect of interparticle interaction in our model using the Density Matrix Renormalization Group and find that when the band is quasi-flat, the ground state is a ferromagnetic Mott insulator. As the band becomes more dispersive, the system undergoes a ferromagnetic to antiferromagnetic transition. We discuss how our model is relevant not only to magic-angle physics in TBG, but also in the larger context of 1D correlations and magnetism.
\end{abstract}

\maketitle

\section{Introduction}

Systems with (quasi)flat bands are ideally suited to hosting novel quantum phases of matter driven by correlation. With kinetic energy suppressed, interparticle interactions are effectively enhanced and decide the fate of the ground state. However, not all (quasi)flat bands are similar in character: they could originate due to different reasons at the single particle level, and, in turn, could result in very different correlation driven ground states. For example, (quasi)flat bands can arise due to weak intersite hopping, geometrically frustrated lattice symmetry, the application of a magnetic field, just to cite a few, with very different ensuing correlated ground states. 

A recent addition in the family of quasi-flat-band systems is twisted bilayer graphene (TBG). When two graphene layers are rotated by a small relative angle away from AA or AB configuration, the low-energy bands become quasi-flat at an angle of $\sim 1^\circ$, called the ``magic angle" \cite{lopes,laissardiere,bistritzer}. A plethora of correlated phases have been seen in experiments \cite{cao1,cao2,dean1,anohall,efetov,koppens,ilani,cao3,arora}. In particular, insulating states that cannot be explained by single particle band theory appear when each of the conduction and valence bands is half-filled. Further, doping away from half-filling gives way to superconducting phases. The phase diagram is reminiscent of that in the cuprates, adding to the excitement, and intense efforts are under way to understand the nature of the ground states. A major challenge in this endeavor is that the single-particle physics of TBG in the quasi-flat-band regime is not well understood. Rotation between the two layers leads to a large scale moir{\'e} pattern and an enlarged supercell which makes \emph{ab initio} calculations impossible \cite{laissardiere}. Long-wavelength descriptions in momentum space, which treat the interlayer coupling perturbatively, exist for general rotations  but break down at the magic angle \cite{bistritzer}. Attempts to write effective real-space lattice theories by constructing appropriate Wannier functions have revealed that the Wannier functions are not very well localized \cite{fu,vafek1,vishwanath1}. Thus, a suitable starting point at the single-particle level is lacking, which has resulted in a variety of competing theories for the correlated physics. Alternative approaches that provide clues to the physics of magic-angle TBG are, therefore, needed.

In this paper, we provide such an approach by defining an auxiliary model which is \emph{a priori} unrelated to TBG, nevertheless captures aspects of TBG both at the single- and many-particle levels. The auxiliary model consists of a two-leg ladder pierced by a flux with the legs coupled by a spatially varying term. We show that at the single-particle level this leads to a quasi-flat low-energy band with characteristics similar to that in magic angle TBG. When interactions are added, at half-filling of the quasi-flat band we find an insulating state with intra- as well as inter-leg ferromagnetic ordering between spins. As the bandwidth increases, both orders turn antiferromagnetic. The auxiliary model is more tractable than the original TBG Hamiltonian, and could, therefore, be useful in unraveling the origin of many aspects of magic angle TBG. Besides, such an approach is useful in discerning which aspects of (quasi)flat-band physics in TBG---single- and many-particle---are general and can be readily extended to other (quasi)flat-band systems and which are specific to TBG depending on the microscopics. In addition, although the model is inspired by TBG, it is relevant in the broader context of 1D correlations and magnetism, independent of TBG. In particular, it serves as a good model Hamiltonian to study the effect of correlations in the quasi-flat-band limit in 1D.  

\begin{figure*}
\centering
\includegraphics[scale=0.50,trim={0 0 0 0},clip]{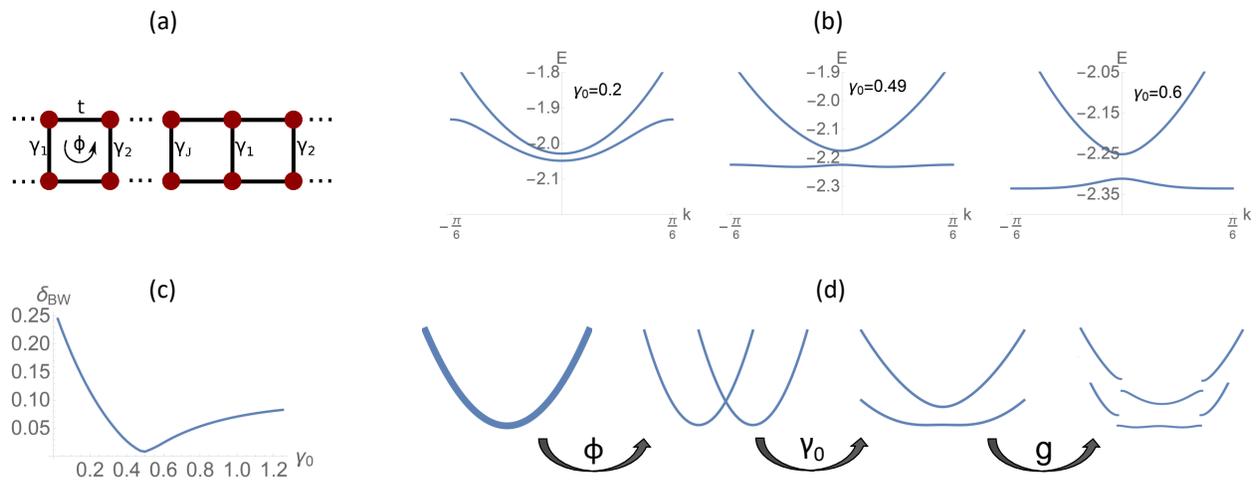}
\caption{(a) Pictorial representation of the auxiliary Hamiltonian defined by Eqs.~(\ref{h12ladder}) and (\ref{hperpladder}) (continuum model) or Eq.~(\ref{intham}) with $U=0$ (lattice model): A two-leg ladder threaded by a magnetic flux and the legs coupled with a spatially varying periodic term. (b) Low-energy band structure of the auxiliary Hamiltonian. Calculations have been performed using Eq.~(\ref{intham}) with $U=0$. Here $t=1$ and $\gamma_j=\gamma_0[1+g\ \mathrm{cos}(2\pi j/J)]e^{i(2\pi j/J)}$ with $g=0.1$ and $J=6$. With increase in $\gamma_0$, the lowest band first becomes quasi-flat and then becomes dispersive. (c) Dependence of the bandwidth of the lowest band on $\gamma_0$. (d) Schematic representation of the origin of the low-energy quasi-flat band---see text for description.}
\label{figband}
\end{figure*}

\section{Model}

The auxiliary model is inspired by the continuum model of TBG. At low energies, a TBG can be described by ($\hbar=1$) \cite{bistritzer,Palarxiv}
\begin{eqnarray}
h_{1,2}&=&-iv_0\begin{pmatrix}
0&\partial_x-i\partial_y\\
\partial_x+i\partial_y&0\label{h12}
\end{pmatrix},\\
h_{\perp}&=&\frac{\gamma_0}{3}\sum_{n=0}^2e^{i \delta\mathbf{K}_n\cdot\mathbf{r}}
\begin{pmatrix}
1&e^{-i \frac{2\pi n}{3}}\\
e^{i \frac{2\pi n}{3}}&1
\end{pmatrix}.\label{hperp}
\end{eqnarray}
where $h_{1,2}$ are Dirac Hamiltonians describing the individual layers and $h_{\perp}$ describes interlayer coupling. Here, $v_0$ is the Dirac velocity of the individual layers, $\gamma_0$ is the interlayer coupling strength, $\delta\mK=\mK^{\theta}-\mK$ is the vector that connects the Dirac points of the rotated and the unrotated layers in momentum space, and $\delta\mK_n$ is $\delta\mK$ rotated by an angle of $2\pi n/3$. The vector $\delta\mb{K}$ sets the length scale of  the moir{\'e} pattern which has a periodicity of $\frac{4\pi}{3\delta K}$. The above Hamiltonian has the interesting property that the bandwidth of the low-energy bands depends on the parameter $\alpha=\frac{\gamma_0}{v_0\delta K}$. As $\alpha$ increases from zero, the bandwidth decreases, approximately vanishes at $\alpha\sim 1$, and increases again with further increase in $\alpha$. While a reduction in the bandwidth can be analytically understood by employing a perturbation theory in $\alpha$, the quasi-flat-band regime can be explored only numerically because perturbation theory breaks down when $\alpha\sim 1$. 

To obtain an understanding of magic-angle TBG, it is imperative we ascertain which aspect(s) of the above Hamiltonian is(are) responsible for the phenomenon of band flattening.
Are Dirac dispersion and band touching necessary? Is it essential for $h_{1,2}$ and $h_{\perp}$ to have a matrix structure (arising from the sublattice degrees of freedom)? Is the phenomenon constrained to only two dimensions? Motivated by these questions, we construct an auxiliary  Hamiltonian by replacing Eqs.~(\ref{h12}) and (\ref{hperp}) with 
\begin{eqnarray}
h_{1,2}&=&-\frac{\partial_x^2}{2m},\label{h12ladder}\\
h_{\perp}&=&\gamma(x)e^{i\phi(x)}\label{hperpladder}.
\end{eqnarray}
The auxiliary Hamiltonian is \emph{a priori} unrelated to the original TBG Hamiltonian but the inspiration is obvious: We have replaced the individual graphene layers with one dimensional free electrons while keeping the coupling term complex similar to the TBG Hamiltonian. We impose the condition $\gamma(x)=\gamma(x+X)$ and $\phi(x)=\phi(x+X)$ so that $h_{\perp}$ is periodic as in TBG. Thus, the auxiliary Hamiltonian is the long-wavelength limit of a two-leg ladder model threaded by a magnetic flux where the legs are coupled by a spatially varying periodic term, as represented in Fig.~\ref{figband}(a) (the corresponding lattice model is given by Eq.~(\ref{intham}) with $U=0$). For simplicity, we assume a constant flux per plaquette and small spatial modulation of $\gamma(x)$: $\phi(x)=2\pi x/X\equiv\phi_0 x$ and $\gamma(x)=\gamma_0[1+g\cos(2\pi x/X)]$, with $g<1$. 

We now show that the resulting low-energy band structure exhibits all the salient features of the bands in TBG. Fig.~\ref{figband}(b) shows that increasing $\gamma_0$, analogous to increasing $\alpha$ in TBG, causes the bandwidth of the lowest-energy band to first decrease, go to a minimum resulting in a quasi-flat band, and then increase again. This is summarized in Fig.~\ref{figband}(c). The mechanism leading to this behavior is traced schematically in Fig.~\ref{figband}(d): First, the complex interleg coupling causes the two degenerate bands to split along the momentum axis by an amount  governed by the complex phase. Next, the constant part of the interleg coupling strength, $\gamma_0$, breaks the degeneracy at the touching point and separates out the low-energy band. Finally, the parameter $g$, describing the spatially varying part of the coupling strength, opens up a gap $\Delta_{BG}$ at the miniband edge due to Bragg scattering and separates out a quasi-flat miniband of bandwidth $\delta_{BW}$. In momentum space, the low-energy bands are described by the truncated Hamiltonian,
\beq
H_{k}=
\begin{pmatrix}
\ve_{k,-1}&\gamma_0&0&0\\
\gamma_0&\ve_{k,0}&g\gamma_0/2&0\\
0&g\gamma_0/2&\ve_{k,0}&\gamma_0\\
0&0&\gamma_0&\ve_{k,+1}
\end{pmatrix},
\label{hk}
\eeq
where $\ve_{k,n}=\frac{(k+n\phi_0)^2}{2m}$. Eq.~(\ref{hk}) reproduces the band structure in Fig.	~\ref{figband} remarkably well, \emph{even in the quasi-flat-band regime} when $\delta_{BW}\ll\Delta_{BG}$, in contrast to magic-angle TBG where this is not possible \cite{bistritzer}---see supplemental material \cite{SuppMaterial}. The band flattening results from a combined effect of the parameters $\gamma_0$ and $g$ (for a fixed $\phi_0$) which can be independently tuned in our model.

Since the flatness depends only on the flux and the spatially varying coupling term, one can replace the two legs of the ladder with any other object with required attributes (e.g., dimensionality, topology, internal degrees of freedom, etc.) which will be conferred on the resulting quasi-flat band(s). This separation of effects can be exploited to design new quasi-flat bands.

\section{Interactions}

We now study the effect of interparticle interaction on the auxiliary Hamiltonian. Several studies have been carried out so far on interacting models of Fermionic ladders with flux, but these explore the regime where the single-particle band is dispersive \cite{giamarchi,narozhny,carr,sela,burrello1,mazza,burrello2,shimshoni}. Here, in contrast, our focus is on the regime where the single-particle band is quasi-flat and isolated from the rest of the dispersive bands with a bandgap. To that end, we write down the corresponding lattice version: 
\begin{equation}
\label{intham}
\begin{split}
H =-&t\sum_{j,\sigma ,\lambda } c_{j+1\sigma \lambda }^{\dagger } c_{j\sigma \lambda } + \sum_{j,\sigma }\gamma_{j} c_{j\sigma 2 }^{\dagger } c_{j\sigma 1 } + \mathrm{h.c.}\\ + & U\sum _{j,\lambda } n_{j \uparrow \lambda } n_{j \downarrow \lambda }.
\end{split}
\end{equation} 
Here, $c_{j\sigma \lambda }$ annihilates an electron on site $j$, on the leg with index $\lambda =1,2$ and spin ${\sigma = \downarrow , \uparrow }$. Hopping between adjacent sites in each leg is parametrized by $t$ and between the two legs by $\gamma_j=\gamma_{0} \left ( 1+g\cos \frac{2\pi j}{J} \right )e^{i\frac{2\pi j}{J}}$, where $\gamma_0$ and $g$ are constants. In our calculation, we choose $J$ to be 6 without any loss of generality, and set $t=1$ to fix the energy scale.
The first line reduces to Eqs.~(\ref{h12ladder}) and (\ref{hperpladder}) in the long-wavelength limit. Interaction is added via an onsite Hubbard interaction of strength $U$ in the second line, where $n_{j \sigma  \lambda }$ is the electron number operator. We are interested in the scenario when the quasi-flat band is half-filled; therefore, we fix the electron density at $n= \frac{1}{2J}$. 

The ground state properties are calculated using the finite-size Density Matrix Renormalization Group (DMRG) method ~\cite{DMRG1,DMRG2,DMRG3} with open boundary conditions. Calculations are performed using the ITensor library~\cite{ITensor}. To minimize finite-size effects, we study systems of various sizes upto $2L=2\times240$ sites with cutoff error less than $10^{-9}$. The energy difference between two sweeps is less than $10^{-7}$ during final sweeps. Total charge- and $S_z$-conservation are implemented for better convergence \cite{SuppMaterial}.

\begin{figure}
\centering
\input{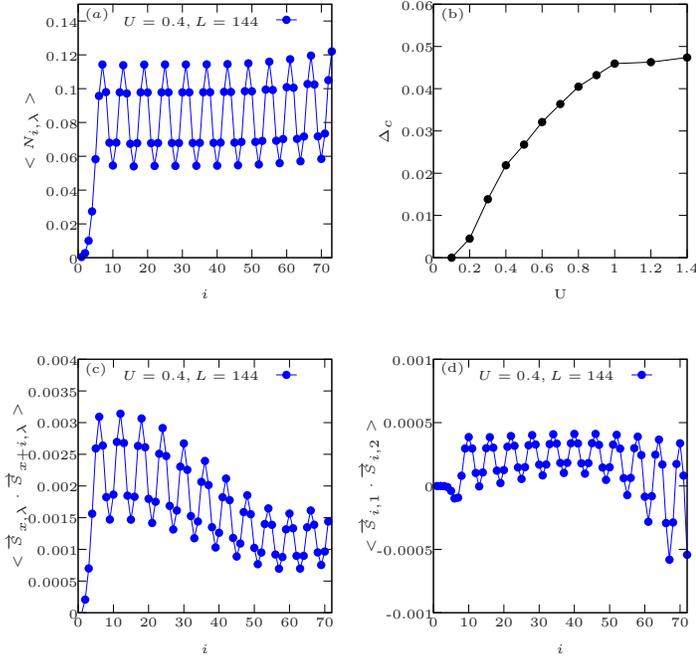}
\caption{\label{Fig2} Ground state properties of the interacting two-leg ladder model described in Eq.~(\ref{intham}) when the lowest noninteracting band is quasi-flat and half-filled. (a) Electron density in real space oscillates with a periodicity of $J=6$. (b) The charge gap at various values of $U$ after finite-size scaling of L: A metal-insulator transition appears at $U\approx 0.1$. 
(c) The intraleg spin-spin correlation. (d) The interleg spin-spin correlation. Correlations in both (c) and (d) are ferromagnetic (only one leg shown in (c); the other leg is similar). Here $x=\frac{L}{4}$ is chosen away from the boundary to minimize finite size effect.}
\end{figure}

In order to stay in the quasi-flat-band regime we fix $\gamma_{0}=0.49$ and $g=0.1$ so that $\delta _{BW}\approx 7.5\times 10^{-3}$ and $\Delta_{BG}\approx 4.89\times 10^{-2}$. For the Hubbard term, we impose the condition $\delta_{BW}\lesssim\langle U\rangle <\Delta_{BG}$, where $\left \langle U \right \rangle = U \sum _{j,\lambda } \left \langle n_{j \uparrow \lambda } n_{j \downarrow \lambda } \right \rangle$ is measured in the non-interacting limit. We choose $U=0.4$ which gives $\left \langle U \right \rangle/\Delta _{BG}\approx 0.18$ and $\left \langle U \right \rangle/\delta _{BW}\approx 1.17$. As shown in Fig.\ref{Fig2}(a), as a result of interaction a commensurate charge density wave emerges with the period of J. 
To identify whether the phase is gapped or not, we calculate the charge gap defined as $\Delta _{c}=\mathrm{\lim}_{L\rightarrow \infty } [ E_{0}\left ( N_{e}= N+2 \right )+E_{0} ( N_{e}=
N-2 )-2E_{0}\left ( N_{e}= N \right )  ]$, where $E_{0}\left ( N_{e} \right )$ refers to the ground state energy of a given electron number $N_e$. 
We find $\Delta_c=0.0214$, obtained after accounting for the finite-size scaling (see supplemental material \cite{SuppMaterial}). The charge gap increases monotonically as $U$ increases, and saturates at $\Delta _{c}=0.0474$ when $U>1$, as shown in Fig.\ref{Fig2}(b). Also, it is seen that $\Delta_c$ goes to zero at $U\approx0.1$, signaling a metal-insulator transition. 
The spin gap is defined similarly as $\Delta _{s}=E_{0}\left ( S_{z}^{tot}=1 \right )-E_{0}\left ( S_{z}^{tot}=0 \right )$ because of the spin U(1) symmetry.
We find that $\Delta_s=0$ (less than $10^{-6}$). The nonzero charge gap and a zero spin gap at finite $U$ is consistent with the expectation that the system at half-filling is a Mott state.

We next investigate the spin-spin correlation in this Mott state. The spin-spin correlation within each leg is defined as $\langle\overrightarrow{S}_{i\lambda }\cdot \overrightarrow{S}_{j\lambda }\rangle$, where $\overrightarrow{S}_{i\lambda }$ is the total spin of the electrons on site $i$ of the $\lambda^\textrm{th} $ leg. As shown in Fig.\ref{Fig2}(c),the spin-spin correlation oscillates with the same period as the electron density.
However, it is not a conventional spin density wave: the spin-spin correlation values are all positive between any two sites, indicating a ferromagnetic ordering. That is, ordering both within a supercell as well as between two supercells is ferromagnetic. A similar behavior is observed for the spin-spin correlation between the two legs, although it is found to be weaker than the intraleg correlation. This is shown in Fig.\ref{Fig2}(d). Note that the ordering is quasi-long range since a true long-range order is forbidden in 1D by the Mermin-Wagner theorem \cite{mermin}. In all, this suggests that in the quasi-flat-band regime, the system is an unusual \emph{ferromagnetic} Mott insulator.  

\begin{figure}
\centering
\input{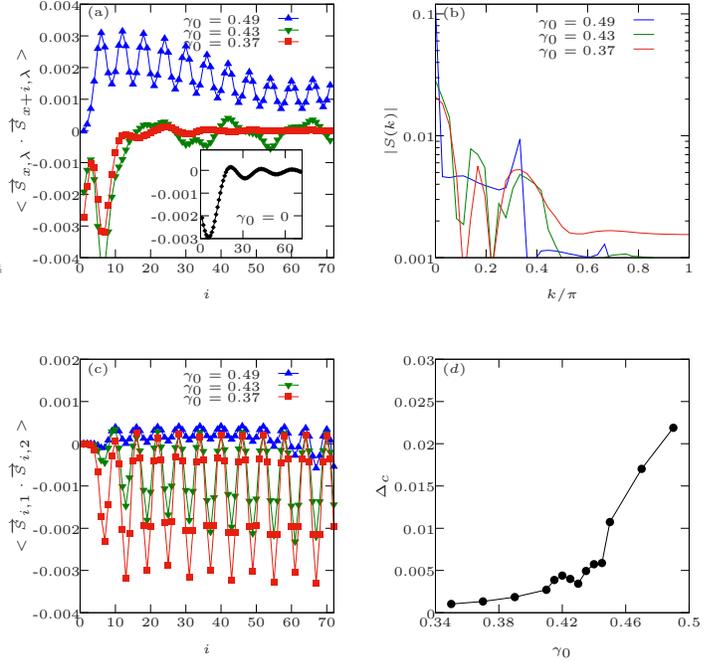}
\caption{\label{Fig3} Effect of increasing the bandwidth by decreasing $\gamma_0$ on the ground state properties shown in Fig.~\ref{Fig2}. (a) The intraleg spin-spin correlation changes from ferromagnetic to antiferromagnetic as $\gamma_0$ decreases (bandwidth increases). The inset shows the correlations in the limit when the two legs are uncoupled. (b) The spin structure factor for various $\gamma_0$. The peaks at $k=0$ are due to the non-zero average value of the correlations. The $k=0$ peak is tallest for the flattest band (largest $\gamma_0$). (c) Same as in (a) but for interleg spin-spin correlation.  (d) The charge gap at various values of $\gamma_{0}$ (bandwidths) after the finite-size scaling. All parameters used, except $\gamma_0$, are the same as those used in Fig.~\ref{Fig2}.}
\end{figure}

It is natural to ask what happens to this phase when the band is no longer quasi-flat. To address this, we keep the interaction strength unchanged but change the bandwidth by changing $\gamma_0$ while keeping $g$ and $\phi_0$ fixed to the values used before. Thus, $\gamma_0$ is a proxy for the bandwidth which can be read off from Fig.~\ref{figband}(c). As shown in Fig.~\ref{Fig3}(a), as $\gamma_0$ is decreased (bandwidth is increased), the intraleg spin-spin correlation changes from ferromagnetic to antiferromagnetic. In the inset, we show the spin-spin correlation in the limit when the two legs are completely uncoupled. Once the transition has occurred, the spin-spin correlation quickly assumes the antiferromagnetic form expected for a single chain. Thus, the emergence of the ferromagnetic phase is contingent on a small bandwidth. 
A better representation of the ferromagnetic--antiferromagnetic transition can be achieved by computing the structure factor $S\left ( k \right ) = \frac{1}{L}\sum_{i,j}\left \langle \overrightarrow{S}_{i,\lambda}\cdot \overrightarrow{S}_{j,\lambda} \right \rangle e^{ik \left ( x_{i} -  x_{j} \right ) }$. We plot in Fig.~\ref{Fig3}(b) $S(k)$ vs. $k$ for different values of $\gamma_0$. Ferromagnetic ordering is indicated by sharp peaks at $k=0$ and $k=2\pi/6=2\pi/J$, which give way to antiferromagnetism at smaller $\gamma_0$ (larger bandwidth) as signaled by a sharp peak at $k=2\pi/12=\pi/J$. Around $\gamma_0\approx 0.43$, where the transition happens, we find both peaks to be sharp. The ferromagnetic-to-antiferromangetic transition is also observed in spin-spin correlation between the two legs as shown in Fig.~\ref{Fig2}(c). Finally, in the charge sector, we find charge density waves with the same periodicity as in the quasi-flat-band regime. The charge gap  decreases with decrease in $\gamma _{0}$ (increase in bandwidth), as shown in Fig.\ref{Fig3}(d), except near $\gamma_0=0.43$ where it shows certain features. This is the same value at which the ferromagnetic-antiferromagnetic transition appears in the spin sector. 

We have confirmed that the above observations do not change qualitatively on introducing an additional nearest neighbor interaction term of the form $V\sum_{j,\sigma,\sigma',\lambda} n_{j\sigma\lambda}n_{j+1\sigma'\lambda}$ in Eq.~(\ref{intham})---see supplemental material \cite{SuppMaterial}.

\section{Discussion}

We now discuss the relevance of these findings in the context of TBG. Experiments have found that near the magic angle, at half-filling of either the conduction or the valence bands, TBG becomes a correlated insulator \cite{cao1,dean1}. Recent theories have proposed that the ground state should be ferromagnetic \cite{vafek2,uchoa}. This has been attributed to the unusual shape of the Wannier functions for the quasi-flat bands and an interplay between the spin and valley degrees of freedom. It is interesting that a ferromagnetic insulating state is also favored in our auxiliary model. Note that while our auxiliary model mimics TBG at the single particle level, the similarity is only in the energetics of the bandwidth, with a completely different microscopic structure---there is no Dirac physics, no valley degree of freedom, and our model is 1D as opposed to 2D---indicating that, perhaps, the underlying physics is quite general. Experimentally, no signature of ferromagnetism has yet been observed at half-filling in TBG. This could be due to the stringent requirement of the narrow bandwidth, so that TBG even slightly away from the magic angle has tendencies toward $\mathbf {k}\neq0$ spin density waves. On the other hand, ferromagnetism has been found recently at three-quarters filling of the conduction band through the observation of anomalous Hall effect \cite{anohall}. It would be interesting, therefore, to extend our model to three quarters and other rational fillings of the band, and also include a valley degree of freedom; the latter can be achieved simply by considering two copies of the two-leg ladder, threaded by flux in opposite directions. Nevertheless, it is important to note that, while such similarities between the original TBG and the auxiliary model provides a novel direction of exploring the physics, there are also limitations to this analogy. For example, critical exponents at the transition point, symmetries of the order parameters, etc., will be different in the two cases due to different dimensions in which the models are defined, and are clearly outside the scope of this analogy. The goal here is to elucidate which aspects of quasi-flat-band physics in TBG can be easily generalized to other (quasi)flat-band systems and which are specific to TBG and its underlying microscopics. 

Beyond TBG, our model is relevant in the larger context of 1D  correlations and magnetism. The emergence of ferromagnetism in the quasi-flat-band regime of our model [Fig.~\ref{Fig2}(a)] seems to contradict the theorem by Lieb and Mattis which states that the ground state in 1D has the lowest possible spin \cite{liebmattis,fazekas}. However, the proof assumes all hoppings and interactions to be real, which is not true in our model. Also, our model should be compared with Tasaki's model \cite{tasaki1,tasakireview} which is known to give rise to flat-band ferromagnetism: Tasaki's model achieves flat bands by having more than one type of atom in the unit cell and beyond-nearest-neighbor-hopping whereas we achieve flat bands with strictly nearest neighbor hopping by including a flux.
Similarly, the metal-insulator (Mott) transition at nonzero $U$ [Fig.~\ref{Fig2}(b)] is at odds with the general result that in 1D such a transition is not expected \cite{1dmott}. The latter is, however, valid only for cases with SU(2) symmetry---it is known to break down in SU(N) generalizations of the 1D Hubbard model \cite{sunhubbard}. Considering the two legs in our model as pseudospins, it is then not surprising that we find a metal-insulator transition at nonzero $U$. Nevertheless, the role of the complex hopping term in this respect cannot be overruled. We believe, by studying our model further, many of these and other similar questions can be answered. Additionally, it would be interesting to conduct a more comprehensive study of the phase diagram using different models of interaction in the quasi-flat-band limit of this model. 

We are not aware of a theory that can adequately describe the model discussed here. A theory explaining the numerical findings will not only provide a basic understanding of many of the observed features in magic-angle TBG, it will also provide a novel direction in the study of 1D correlations and magnetism. These ideas can be experimentally tested independently using cold atoms where Fermionic flux ladders can be simulated \cite{coldatom1}.

\section{Concluding remarks}

In summary, we have shown that a two-leg ladder threaded by a flux with the legs coupled by a spatially varying periodic term produces a low-energy quasi-flat band with characteristics similar to that in magic angle TBG. In the presence of interactions, the ground state is a ferromagnetic Mott insulator, which becomes antiferromagnetic as the band becomes more dispersive. Further study of this model can be helpful in understanding other aspects of magic-angle TBG---by comparing results obtained from the auxiliary and the original models, one can conclude which aspects of TBG are general, independent of microscopics and which are not. On the other hand, independent of TBG, our model is relevant in the larger context of 1D correlations and magnetism. It can serve as a good starting point to explore effects of interaction in 1D when the underlying band is quasi-flat.

\begin{acknowledgments}
Y.H. would like to thank Shun-Chiao Chang for helpful discussions on the DMRG implementation and acknowledge support by the Texas Center for Superconductivity and the Welch Foundation Grant No. E-1146. Numerical calculations was completed in part with resources provided by the Center for Advanced Computing and Data Science at the University of Houston. P. H. would like to thank the Department of Physics, the College of Natural Sciences and Mathematics and the Division of Research at the University of Houston for support. H.K.P. would like to thank Ashvin Vishwanath, Ganapathy Baskaran, Subir Sachdev, Soumya Bera, and  Sumiran Pujari for helpful discussions, and  IRCC, IIT Bombay for financial support via grant RD/0518-IRCCSH0-029.
\end{acknowledgments}


\begin{widetext}

\section{Supplemental Material}

\subsection{The auxiliary Hamiltonian truncated in momentum space}

The noninteracting auxiliary Hamiltonian is defined as (ignoring spin)
\begin{equation}
H=-t\sum_j(c_{j+1,1}^\dagger c_{j,1}+\mathrm{h.c.})-t\sum_j(c_{j+1,2}^\dagger c_{j,2}+\mathrm{h.c.})+\sum_j(\gamma_jc_{j,2}^\dagger c_{j,1}+\mathrm{h.c.}),\label{auxhamsupp}
\end{equation}
where $\gamma_j=\gamma_0(1+g\mathrm{cos} \phi_j)e^{i\phi_j}$, with $\phi_j=\frac{2\pi}{J} j=\phi_0 j$. We make the following gauge transformations: $c_{j,1}\rightarrow c_{j,1}$ and $c_{j,2}\rightarrow e^{i\phi_j}c_{j,2}$, and then carry out a Fourier transform: $c_{j,\alpha}=\frac{1}{\sqrt{J}}\sum_ke^{ikj}c_{k,\alpha}$. Eq.~(\ref{auxhamsupp}) then becomes
\begin{equation}
H=-2t\sum_k \mathrm{cos}\left(k\right)c_{k,1}^\dagger c_{k,1}-2t\sum_k \mathrm{cos}\left(k+\phi_0\right)c_{k,2}^\dagger c_{k,2}
+\gamma_0\left(1+g\mathrm{cos}\phi_0\right)\sum_k\left(c_{k,2}^\dagger c_{k,1}+\mathrm{h.c.}\right).
\end{equation}
Since we are interested only in the low-energy spectrum for $k\in [-\phi_0/2,\phi_0/2]$, we can truncate the Hamiltonian in momentum space and write $H=\sum H_k$ with
\begin{equation}
H_k=
\begin{pmatrix}
c_{k-\phi_0,1}^\dagger&c_{k-\phi_0,2}^\dagger&c_{k,1}^\dagger&c_{k,2}^\dagger
\end{pmatrix}
\cdot
\begin{pmatrix}
\varepsilon_{k,-1}&\gamma_0&0&0\\
\gamma_0&\varepsilon_{k,0}&\gamma_0g/2&0\\
0&\gamma_0g/2&\varepsilon_{k,0}&\gamma_0\\
0&0&\gamma_0&\varepsilon_{k,+1}
\end{pmatrix}
\cdot
\begin{pmatrix}
c_{k-\phi_0,1}\\c_{k-\phi_0,2}\\c_{k,1}\\c_{k,2}
\end{pmatrix},
\label{hksupp}
\end{equation}
where $\varepsilon_{k,n}=-2t\mathrm{cos}(k+n\phi_0)$ which in the long-wavelength limit becomes $\frac{(k+n\phi_0)^2}{2m}$ with $m=\frac{1}{2t}$ (discarding the constant shift of energy equal to $-2t$). As seen in Fig.~\ref{bandsupp}, Eq.~(\ref{hksupp}) reproduces the band structure obtained from Eq.~(\ref{auxhamsupp}) remarkably well, even in the quasi-flat-band regime. 

\begin{figure}[b]
\centering
\subfigure[]{\includegraphics[width=0.3\columnwidth]{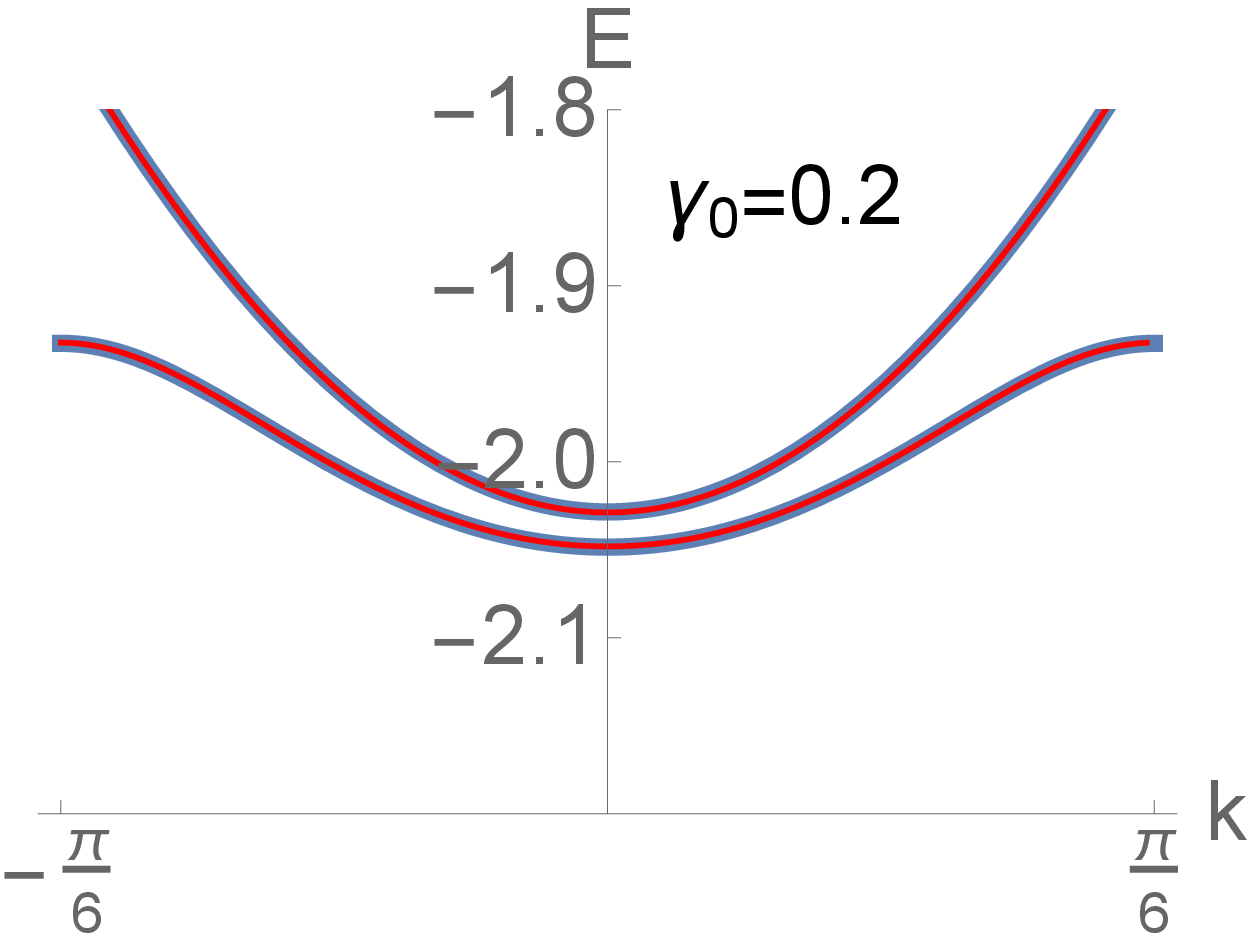}
\label{fig2a}}
\quad
\subfigure[]{\includegraphics[width=0.3\columnwidth]{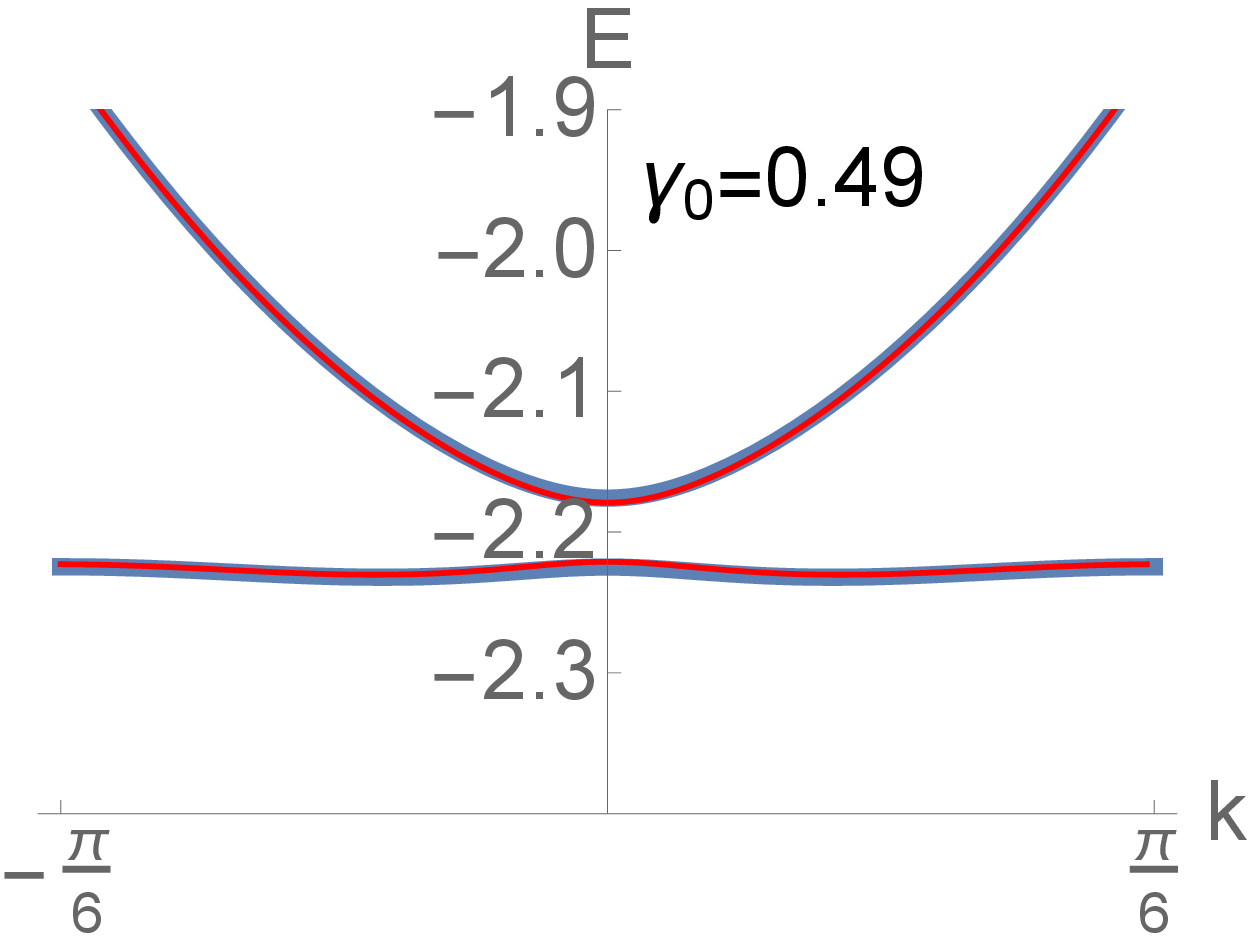}
\label{fig2b}}
\quad
\subfigure[]{\includegraphics[width=0.3\columnwidth]{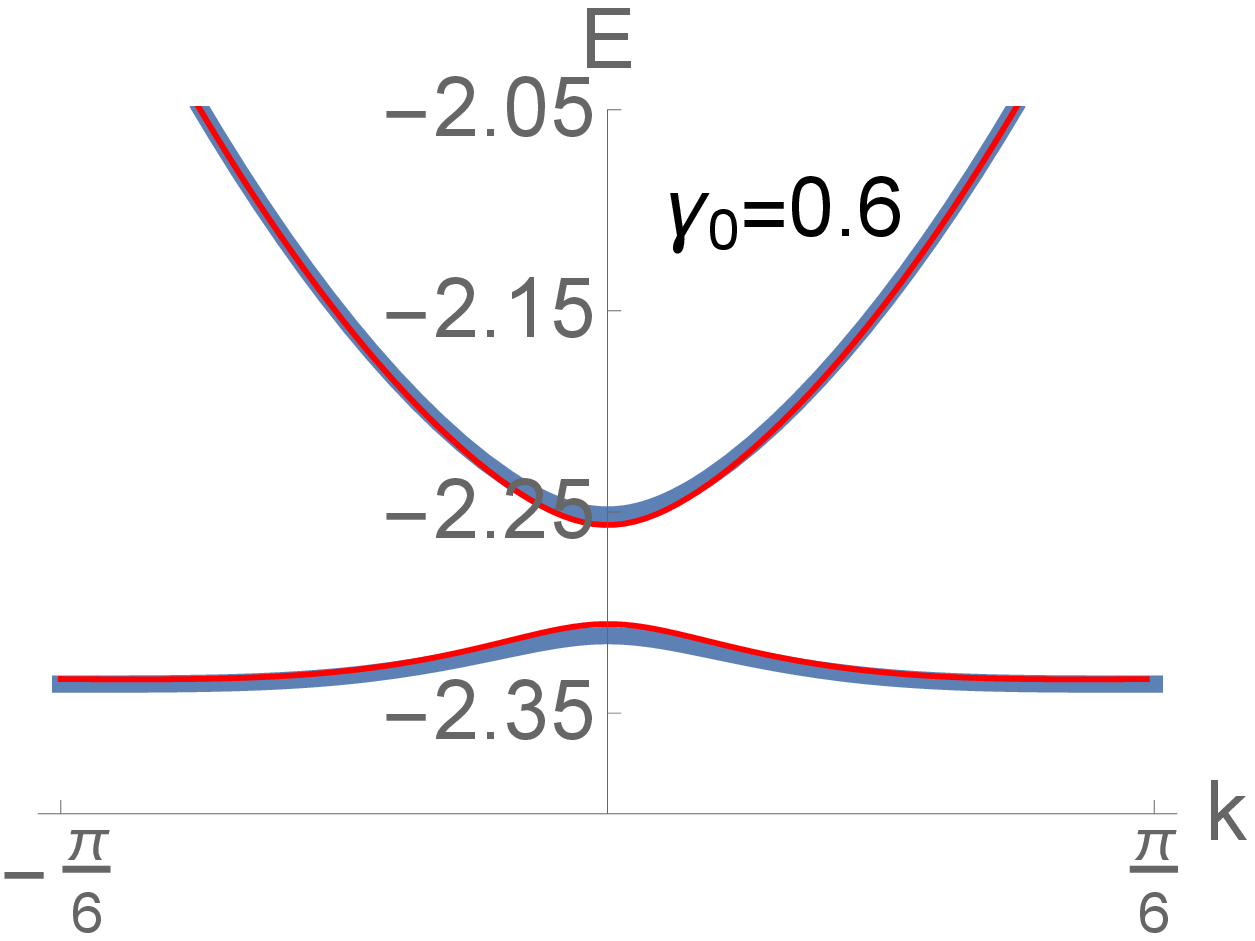}
\label{fig2c}}
\caption{Comparison of the low-energy bands of the auxiliary Hamiltonian derived from exact tight-binding calculation (blue) and from the truncated Hamiltonain [Eq.~(\ref{hksupp})] (red). Here, $t=1$, $\phi_0=2\pi/6$, and  $g=0.1$.}
\label{bandsupp}
\end{figure}

\subsection{The convergence of numerical results}

The convergence of the Density Matrix Renormalization Group (DMRG) results can be checked by the truncation error and ground state energy with increased number of states kept. As shown in Fig.\ref{FigS1}, the ground state energy $E_{0}$ remains almost unchanged as the number of states increases, indicating that the numerical results are converged. Meanwhile the local spin value $\left \langle S^{z}_{i} \right \rangle$ remains zero with an error bar in the order of the truncation error for any finite $U$.

\begin{figure}
\centering
\input{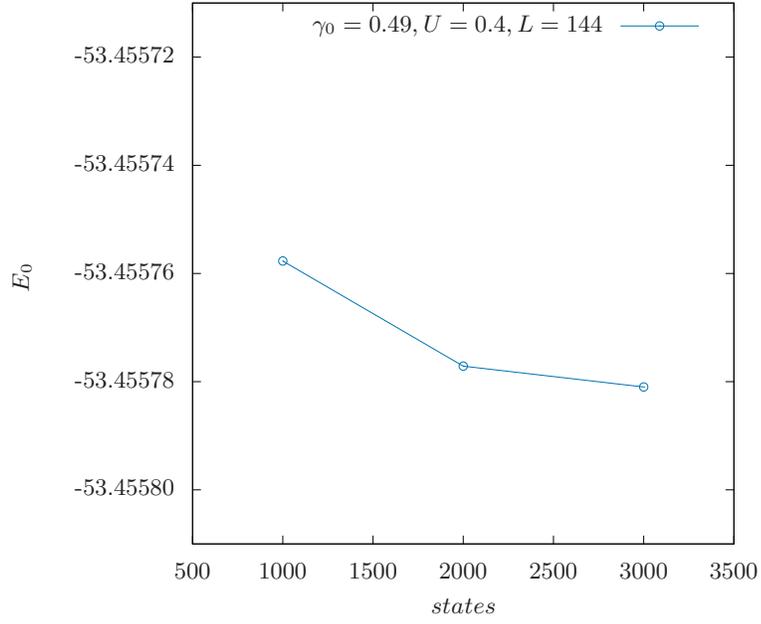}
\caption{\label{FigS1}The groud state energy with various number of states kept. The truncation error is $1 \times 10^{-9}$ for 1000 states, $1 \times 10^{-10}$ for 2000 states, and $2 \times 10^{-11}$ for 3000 states.}
\end{figure}

\subsection{The finite-size scaling of the charge gap}

The charge gap is defined as $\Delta _{c}(L)=E_{0}\left ( N_{e}= N+2 \right )+E_{0} ( N_{e}=N-2 )-2E_{0}\left ( N_{e}= N \right ) $. The calculations of the charge gap depend only on the ground state energy; thus, they are very reliable. Fig.\ref{FigS3} shows the finite-size scaling of the charge gap. The charge gap remains finite after the scaling for $U > 0.2$, which indicates a gapped phase.

\begin{figure}
\centering
\input{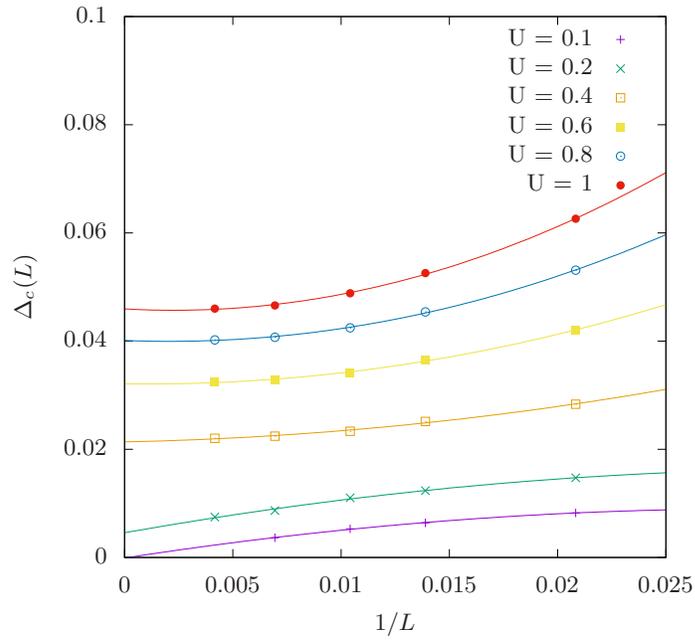}
\caption{\label{FigS3}The finite-size scaling of the charge gap for various U in the quasi-flat-band regime. We have used a least-square fit to the third order of polynomials in 1/L. Similar fittings are used in the extrapolation of the charge gap for various $\gamma_{0}$.}
\end{figure}

\subsection{The spin correlations under the next-nearest-neighbor interactions}

Besides the onsite Hubbard interactions, we have tested other interactions such as the next-nearest-neighbor Coulomb interactions, which is defined as $V\sum_{j,\sigma ,{\sigma}' , \lambda } n_{j\sigma \lambda } n_{j+1{\sigma}' \lambda }$. As shown in Fig.\ref{FigS4} the spin correlations become ferromagnetic for finite V, which is similar to the one with only onsite Coulomb interactions, suggesting that the $ferromagnetic$ Mott state is robust against various types of Coulomb interactions.

\begin{figure}
\centering
\input{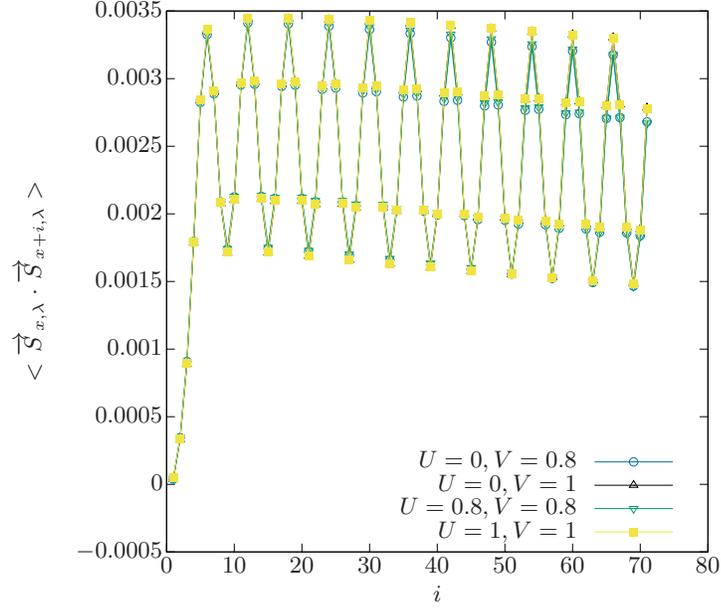}
\caption{\label{FigS4}The intra-chain spin correlations with next-nearest-neighbor Coulomb interactions, obtained for the system of length $L= 144$. We choose $x=\frac{L}{4}$ in order to minimize the effect of the open boundary. Only one leg is shown, the other is the same.}
\end{figure}

\end{widetext}

\end{document}